\documentclass[3pt,onecolumn]{article}
\usepackage{graphicx}

\newcommand{\newsection}[1]{
\addtocounter{section}{1}
\setcounter{equation}{0}
\setcounter{subsection}{0}
\addcontentsline{toc}{section}{\protect
\numberline{\arabic{section}}{{\rm #1}}}
\vglue .6cm
\pagebreak[3]
\noindent{\bf  \thesection. #1}\nopagebreak[4]\par\vskip .3cm}

\newcommand{\ben}{\begin{enumerate}}
\newcommand{\een}{\end{enumerate}}

\newcommand{\be}{\begin{equation}}
\newcommand{\ee}{\end{equation}}
\newcommand{\bea}{\begin{eqnarray}}
\newcommand{\eea}{\end{eqnarray}}

\begin{document}
\begin{flushright}
\end{flushright}
\vspace{0.1cm}
\thispagestyle{empty}

\begin{center}
{\Large\bf 
Twisting Uneven Ropes
}\\[10mm]
{\rm  Kasper W. Olsen\footnote{kaspeolsen@math.ku.dk}\\
Department of Mathematical Sciences, University of Copenhagen\\ 
Universitetsparken 5, 2100 Copenhagen \O, Denmark}\\[10mm]
\end{center}

\addtocounter{section}{1}

A classical two-stranded rope can be made by twisting two identical strands together under strain \cite{bohr2011}.
Despite being conceptually simple, the contact-equations for helically twisted identical strands have only 
been solved within the last 20 years \cite{pieranski2001,neukirch2002}. For a bundle of strands, it is convenient to 
use the metric equivalence between filament packing in twisted bundles and disk packing on a non-Euclidean surface \cite{grason2012}. 
Our goal here is modest -- to understand the twisting of two circular strands, where one is thicker than the other. This is what we call an uneven rope. Examples of uneven rope have been briefly addressed in the literature, for example in \cite{mayrhofer2012,starostin2014}.
A variant of the uneven rope is the wire ropes where the core typically has a different diameter than the helically wound outer layer and therefore is made of wires of a different diameter such that they fit together \cite{costello1997}. Another variant is the viking neck and arm rings, which are made of (gold and silver) wires twisted together with thinner wires \cite{olsen2010a}.

We let the diameter of one strand be denoted by $D$ and the other one by $d = (1-\epsilon) D$, where $0\leq \epsilon < 1$ is a deformation parameter.
First, we assume that the center lines of the rope form simple helices, as in \cite{bohr2011} (in practice it might be difficult to twist the strands in this way since one is thicker than the other and hence has a higher bending rigidity). The D-strand centreline is twisting around a cylinder of radius $a$ and pitch $H=2\pi h$, hence its parametrisation is $ {\bf r} = (a\cos t, a \sin t, h t) ,$
with the pitch angle (angle with the $x-y$ plane) determined by 

\be
\tan v_\bot = \frac{h}{a}
\ee
The d-strand centreline is twisted on a cylinder of radius $a^*$, i.e. its parametrisation is ${\bf r}^* = (a^*\cos t, a^* \sin t, h^* t+\delta)$.
Since the two strands are twisted on the same cylinder, we have $a-a^* = \epsilon D/2$, $h^*=h$, and 
$\delta = \pi h$.
 The corresponding d-strand pitch angle $v_\bot^*$ must be given by 
\be
\tan v_\bot^* = \frac{h}{a^*} = \frac{\tan v_\bot}{1-\epsilon\frac{D}{2a}}
\ee
The ratio between cylinder diameter and strand diameter, $2a/D$, is a function of both $v_\bot$ and $\epsilon$ and its determination is described in the appendix. However, it should be obvious that $2a/D\rightarrow 1$ for $v_\bot \rightarrow 90^\circ$ (parallel vertical strands), and that $2a/D\rightarrow \infty$ for $v_\bot \rightarrow 0^\circ$ (parallel horizontal strands).

The length $L_r$ of a rope obtained by twisting two strands of length $L_s$ must be

\begin{equation}
L_r = L_s\sin v_\bot,
\end{equation}
and the number of turns $n_r$ on the two-stranded rope is
\begin{equation}
n_r = \frac{L_s}{2\pi a}\cos v_\bot\, .
\end{equation}
Solving these two equations for a classical rope gives rise to the so-called  {\it rope curve} \cite{bohr2011}. This curve display $L_r$ as a function of $n_r$ with the apex corresponding to the maximally twisted geometry: at this pitch angle the number of turns is maximal and hence the rope will not twist one, or the other way, under strain (in the context of wire rope, this is called a rotation-resistant rope). For a classical two-stranded 
rope ($\epsilon=0$) the corresponding zero-twist pitch angle was shown to be $v_{ZT} = 39.4^\circ$ \cite{olsen2011}.

Let us return to the uneven rope. For such a rope, we see from eq. (1.2) that $v_\bot^* \geq v_\bot$, and hence if twisted from two strands of even length, we will have $L_r^* \geq L_r$, where $L_r^*$ is the length of the twisted d-strand. Figure 1 show the d-strand pitch angle $v_\bot^*$ as a function of D-strand pitch angle $v_\bot$, as well as the ratio $L_r^*/L_r$. The parameter $\epsilon$ is varied from $\epsilon = 0.0$ to $\epsilon = 0.5$ in discrete steps. 

\begin{figure}[h]\centering
\includegraphics[width=6cm]{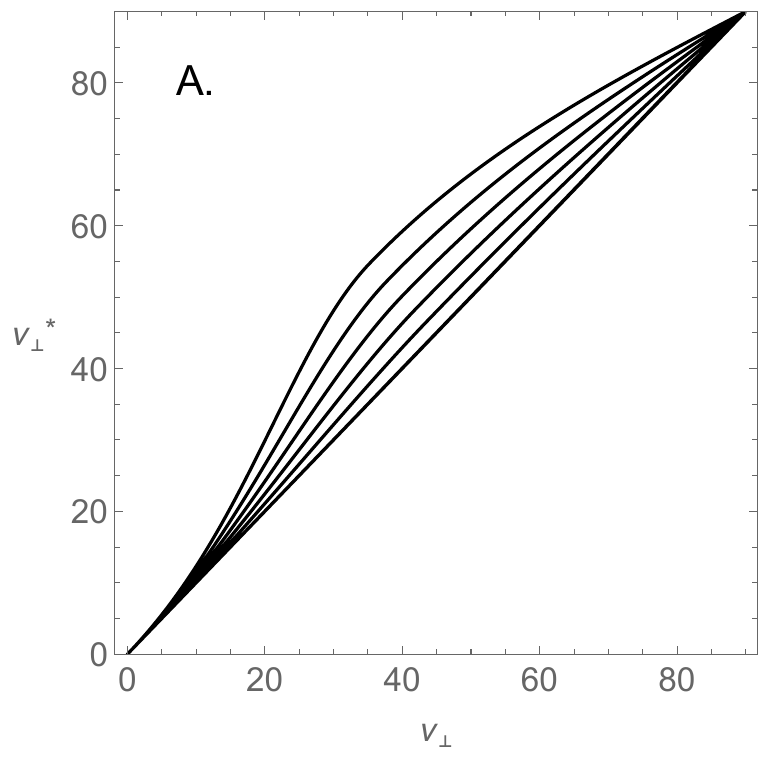}
\includegraphics[width=6.1cm]{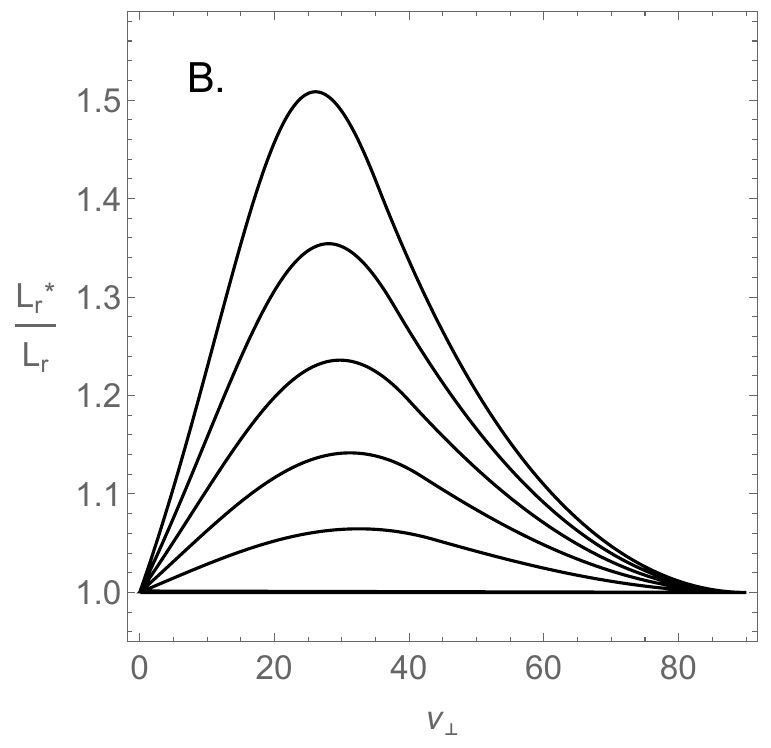}
\caption{\it (A) d-strand pitch angle $v_\bot^*$ as a function of D-strand pitch angle $v_\bot$. The parameter $\epsilon$ is varied from $\epsilon = 0.0$ to $\epsilon = 0.5$ in discrete steps of length $\Delta\epsilon=0.1$. For example, with $\epsilon =0.5$ we have $v_\bot^* = 63.4^\circ$ when $v_\bot = 45^\circ$; (B) the length-ratio $L_r^*/L_r$ between the twisted d-strand and D-strand. The location of the maximum approaches $33.7^\circ$  for small $\epsilon > 0$.}
\end{figure}

From fig. 1B it is clear that a working rope cannot be constructed by using strands of the same length, but different diameters. Generally, one strand will be appear longer in the end of the process of laying the rope. For example, if we want a working rope with $v_\bot = 45^\circ$ then the D-strand should be $\sim 1.265$ times longer than the d-strand if $\epsilon = 0.5$ (fig. 1B).

Now we can discuss the rope curve for the uneven rope: let us assume that the two strands both have total length $L_s$. Then for a given D-strand diameter and deformation parameter $\epsilon$, we can use Eqs. (1.3) and (1.4) to determine the length of the rope, $L_r(n_r)$ as function of the number of turns. Obviously, we can write

\begin{equation}
n_r = \frac{L_s}{\pi D}\frac{D}{2a}\cos v_\bot\, ,
\end{equation}

\noindent hence $n_r$ depends on $\epsilon$ through the term $D/2a$. In fig. 2A we have plotted the rope-length for $L_s=1$, $D= 1/100$ and $\epsilon=0.0$ to $0.5$ in discrete steps of length $\Delta\epsilon=0.1$. Along the solid curve, the two strands are in contact. The upper part of the curves correspond to strands twisted together under strain, beginning with straight strands placed next to one another.
The maximum number of turns is seen to be $\sim 23.5$ when $\epsilon = 0$. Obviously, the maximum number of turns of the rope increases as the d-strand become thinner. But the zero-twist pitch angle also decreases, see Table 1 (page 4) and the plot in fig. 2B.  The lower part of the curves correspond to strands twisted on a virtual cylinder of decreasing diameter.

\begin{figure}[h]\centering
\includegraphics[width=6.4cm]{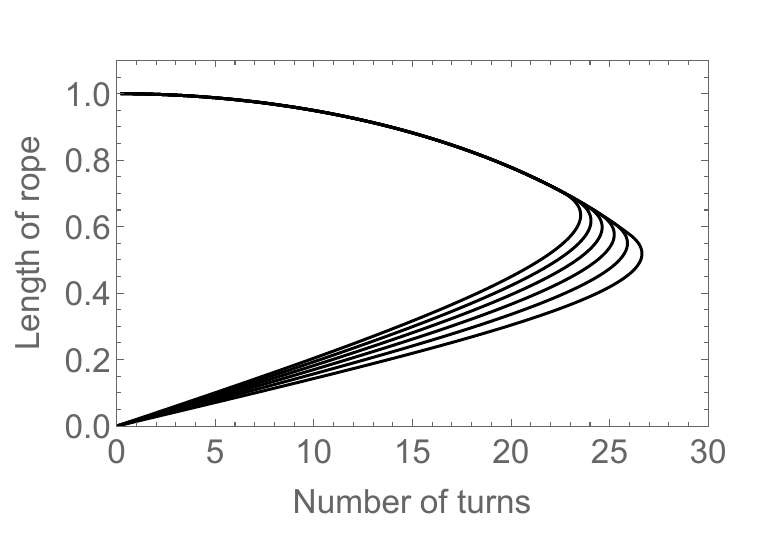}
\hspace{.5cm}
\includegraphics[width=6cm]{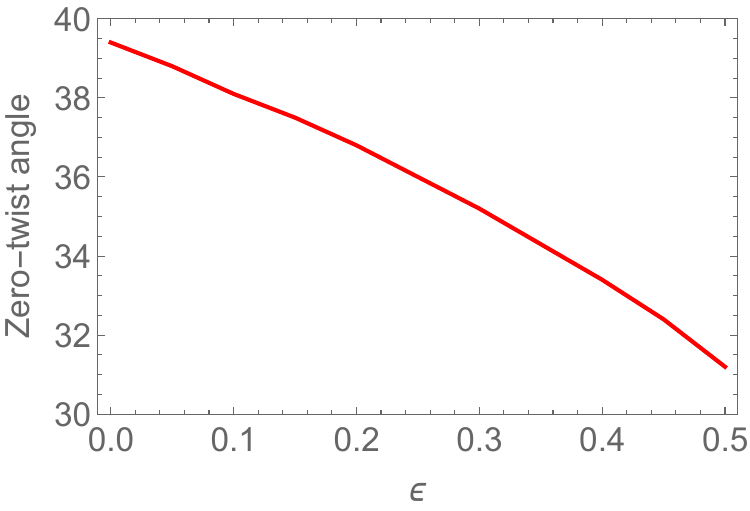}
\caption{\it (A) The rope curve, $L_r(n_r)$, for $\epsilon = 0.0, 0.1,\ldots , 0.5$ with $\epsilon$ increasing to the right. The length of the rope at the apex point (zero-twist) decreases as a function of $\epsilon$. (B) The corresponding zero-twist pitch angle, $v_{ZT}$, plotted as a function of $\epsilon$.}
\end{figure}
After figuring out the rope curve for an uneven rope, we can put some past findings into context.
In practice, a classical rope made of two "identical strands" will be twisted from strands that do not have the exact same diameter. If e.g. $\epsilon=0.01$ then we find $v_{ZT}= 39.3^\circ$ which is $0.2\%$ away from the "classical" zero-twist angle of $39.4^\circ$ \cite{olsen2011}.  

The rope curve for an uneven rope might also explain how a {\it single} length of wire could be twisted in repetitive and beautiful patterns, with rather small pitch angles, as in neck and arm rings produced in the Bronze Age (e.g. torcs, see \cite{rahmstorf2019}). 
For a single circular wire, there is no natural zero-twist pitch angle \cite{bohr2011}. A "help-wire" (d-strand) could be twisted together with the thicker wire, at or close to the zero-twist structure determined by the value of $\epsilon = 1-d/D$, and then discarded in the end. 
It is unknown whether such a design has been used in nature, for example in the context of biomolecules. 
Figure 3 depicts the uneven rope ($\epsilon = 0.0, 0.2$, and $0.4$) with a pitch angle of the respective zero-twist structures.

A generalisation of the uneven rope to a higher number of strands is certainly feasible. In the context of mapping to the metric equivalent non-Euclidean surface, it would correspond to disc packings of variable geodesic diameter, see e.g. \cite{grason2012,grason2014}. Another issue worth studying is the stability of the solutions corresponding to uneven rope using an energy functional, which will depend on material properties of the strands \cite{mahadevan2019,borum2020}.

\begin{figure}[h]\centering
\includegraphics[width=2.1cm]{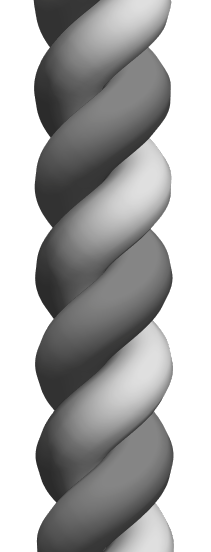}
\hspace{1cm}
\includegraphics[width=2.0cm]{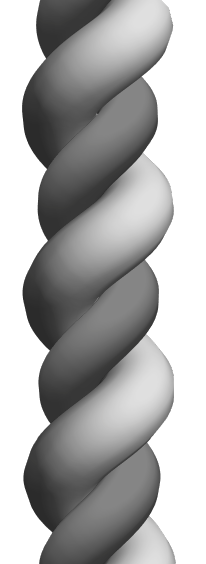}
\hspace{1.2cm}
\includegraphics[width=1.8cm]{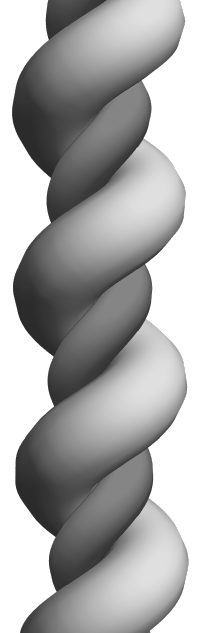}
\caption{\it The maximally twisted ropes (ZT geometries) for deformation parameter $\epsilon = 0.0, 0.2, 0.4$ (from left to right); the ratio between the strand diameters is $d/D=1-\epsilon$.}
\end{figure}


\begin{table}[h]
\caption{Zero-twist pitch angle (deg.) as a function of deformation parameter $\epsilon$; the location of the maximum
in fig. 1B is denoted by $v_\bot^*$.}
\begin{center}
\begin{tabular}{ccccccccccccc}
\hline
$\epsilon$ & 0.0 & 0.1 & 0.2 &  0.3 &   0.4 & 0.5 \\
$v_{ZT}$  & 39.4  & 38.1   & 36.8   & 35.2   & 33.4   & 31.2\\
\hline
$v_{\bot}^*$ & - & 32.5 & 31.2 & 29.7 & 28.0 & 26.1\\
\hline
\end{tabular}
\end{center}
\label{default}
\end{table}%

\section*{Acknowledgments}

I thank Greg Grason for correspondence on the geometry of twisted filaments, and Carsten Hvid at the Viking Ship Museum for discussion on the practicality of twisting uneven ropes.

\newpage
\appendix
\newsection{Appendix}

The distance squared between two points $(t_1,t_2)$ on the D- and d-strand is

\begin{eqnarray*}
\Delta^2(t_1,t_2) &=& (a^*\cos t_2-a\cos t_1)^2 + (a^*\sin t_2-a\sin t_1)^2 +(h^*t_2+\delta -h t_1)^2\\
&=& 2a^2-\epsilon aD+\epsilon^2D^2/4-(2a^2 -\epsilon aD)\cos(t_2-t_1)\\
& & + (ht_2+\delta -h t_1)^2,
\end{eqnarray*}
where $a^*=a-\epsilon D/2$ and $h^*=h$.
Since this distance only depends on $t:=t_2-t_1$, we have

\begin{eqnarray*}
\Delta^2(t) 
&=& 2a^2 -\epsilon aD+\epsilon^2D^2/4-(2a^2-\epsilon aD)\cos t\\
& & + (h t+\delta)^2, 
\end{eqnarray*}
with $\delta = \pi h$.
For twisting ropes, two conditions have to hold: (1) that the distance between the two ropes satisfy $ \Delta = D/2+(1-\epsilon) D/2$ and (2) the derivative of $\Delta^2(t)$ should be zero, where

\be
\frac{d}{dt}\Delta^2 = \sin t (2a^2-\epsilon aD)+2h^2t+2\pi h^2\, .
\ee
Therefore, 

\begin{equation}
(1-\epsilon\frac{D}{2a})\sin t+\frac{h^2}{a^2}t + \frac{h^2}{a^2}\pi = 0\, ,
\end{equation}
and

\begin{eqnarray*}
D^2(1-\epsilon/2)^2
&=& 2a^2-\epsilon aD+\epsilon^2D^2/4-(2a^2-\epsilon aD)\cos t\\
& & + (h t+\delta)^2, 
\end{eqnarray*}
i.e. 

\begin{equation}
\frac{D}{2a}(1-\epsilon/2) = \frac{1}{\sqrt{2}}
\sqrt{
1 -\epsilon\frac{D}{2a}+\frac{\epsilon^2}{2}\left(\frac{D}{2a}\right)^2-(1-\epsilon \frac{D}{2a})\cos t
 + \frac{1}{2}(h t/a+h\pi/a)^2 
 }\, . 
\end{equation}

For $\epsilon=0$, these equations have been solved numerically in \cite{olsen2009}: given $h/a$, equation (A.2) is solved for $t$ corresponding to packing between strands, and then this $t$ is inserted in (A.3) to determine $2a/D$.
Here, with $\epsilon \neq 0$, given ($\epsilon, h/a$) we solve the two equations (A.2) and (A.3) numerically for the two unknowns $(t, D/2a)$.

In the usual two-stranded rope, the geometry is not limited by bending effects, meaning that all pitch angles are possible \cite{olsen2009}. This might be different for the uneven rope.
The usual curvature condition that prohibits self-intersection is \cite{olsen2009}
\be
D \leq \frac{2}{\kappa} \Rightarrow \frac{2a}{D} \geq a\kappa
\ee
where $\kappa$ is the curvature of the D-strand centreline. We have for a simple helix,  $a \kappa =  a^2 \left( a^2 +h^2\right)^{-1} = \cos^2v_\bot$. For the $d$-strand the condition becomes

\be
(1-\epsilon)D \leq \frac{2}{\kappa^*} \, ,
\ee
where 
$a^* \kappa^* = \left( 1+(h^*/a^*)^2\right)^{-1}$, and $\kappa^*$ is the curvature of the d-strand centreline. Therefore, this condition can be rewritten as

\be
\frac{2a}{D} \geq (1-\epsilon) a\kappa^* = \frac{(1-\epsilon) 2a(2a-\epsilon D)}{(-2a+\epsilon D)^2+4h^2}
= \frac{(1-\epsilon) (1-\epsilon D/2a)}{(1-\epsilon D/2a)^2+(h/a)^2}
\ee
Figure 4 show $2a/D$ as a function of $v_\bot$ (blue), and the curves limiting bending, i.e. the minimal allowed values for $2a/D$ (black). It is observed that these different curves do not intersect, hence there is no self-intersection of the uneven ropes.

\begin{figure}[h]\centering
\includegraphics[width=6cm]{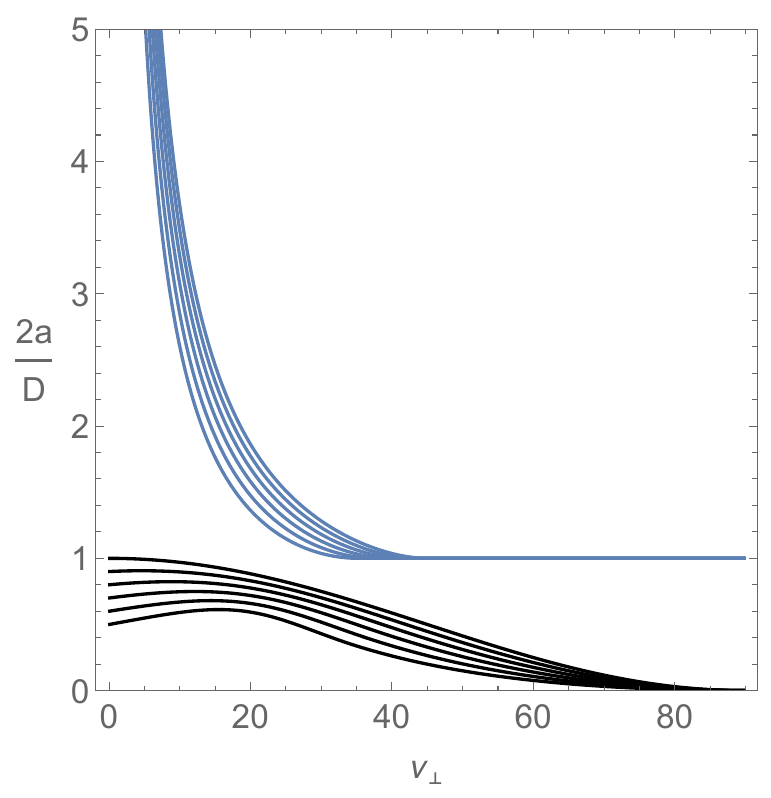}
\caption{\it The figure show the ratio $2a/D$ (blue curve) as a function of pitch angle $v_\bot$. The lower (black) curves is the rhs of eq. (A.6). The parameter $\epsilon$ is varied from $\epsilon = 0.0$ to $\epsilon = 0.5$ in discrete steps of length $\Delta\epsilon=0.1$ with $\epsilon$ increasing from above on the plot. 
}
\end{figure}

For completeness, we also calculate the volume fraction of the uneven rope. The volume fraction, $f_V$,  can readily be defined for the uneven rope as it was introduced for a double helix in \cite{olsen2009}. We define $f_V = V_S/V_E$, where $V_S$ is the volume of the two uneven strands and $V_E$ is the volume of an enclosing cylinder. This cylinder has height $2\pi h$ and diameter $2a+D$. We find the following expression for $f_V$:

\be
f_V =\left( \frac{2a}{D}+1\right)^{-2}\left[
\left( \frac{a^2}{h^2}+1\right)^{1/2}
+(1-\epsilon)^2\left( \frac{a^2}{h^2}\left( 1-\epsilon\frac{D}{2a}\right)^2+1\right)^{1/2}\right]
\ee
Table 2 show the resulting pitch angle corresponding to optimal packing (CP) as a function of $\epsilon$.

\begin{table}[htp]
\caption{Close-packed pitch angle (deg.) and volume fraction as a function of deformation parameter $\epsilon$}
\begin{center}
\begin{tabular}{cccccccc}
\hline
$\epsilon$ & 0.0 & 0.1 & 0.2 & 0.3 & 0.4 & 0.5 & 0.51 \\
$v_{CP}$ & 32.4 & 29.7 & 27.0 & 24.2 & 21.7 & 19.4 & 19.2 \\
$f_V$ & 0.769 & 0.707 & 0.659 & 0.626 & 0.610 & 0.614  & 0.615\\
\hline
\end{tabular}
\end{center}
\label{default}
\end{table}%

\end{document}